\documentclass[preprint,showpacs,showkeys,aps]{revtex4}

\usepackage{graphicx}

\begin{document}

\title{Local Gram-Schmidt and Covariant Lyapunov Vectors and Exponents for
Three Harmonic Oscillator Problems}
\author{
Wm. G. Hoover and
Carol G. Hoover                                              \\
Ruby Valley Research Institute                               \\
Ruby Valley, Nevada 89833                                    \\
}

\date{\today}



\begin{abstract}

We compare the Gram-Schmidt and covariant phase-space-basis-vector
descriptions for three time-reversible harmonic oscillator problems,
in two, three, and four phase-space dimensions respectively. The 
two-dimensional  problem can be solved analytically.  The
three-dimensional and four-dimensional problems studied here are
simultaneously chaotic, time-reversible, and dissipative.  Our
treatment is intended to be pedagogical, for July 2011 publication in
Communications in Nonlinear Science and Numerical Simulation,
and for use in the second edition of our book on {\it Time Reversibility,
Computer Simulation, and Chaos}.  Comments are very welcome.

\end{abstract}

\maketitle

\section{Introduction}

It is Lyapunov instability which makes statistical mechanics possible\cite{b1}.
For stationary boundary conditions, either equilibrium or nonequilibrium,
the {\it exponential growth}, $\propto e^{\lambda t}$, of small phase-space
perturbations
$\{ \delta q,\delta p\}$ or $\{ \delta q,\delta p,\delta \zeta \}$,
``sensitive dependence on initial conditions'', provides longtime
averages independent of initial conditions.  In addition to the
coordinates and momenta $\{ \ q,p \ \}$, time-reversible
``thermostat variables'' $\{ \ \zeta \ \}$ can be used to impose
{\em nonequilibrium} boundary conditions\cite{b2}, such as velocity or
temperature gradients.  A prototypical  nonequilibrium problem simulates
heat flow between two thermal reservoirs maintained at temperatures $T_H$
and $T_C$.  The resulting heat flow through an internal  Newtonian region,
bounded by the two reservoirs, can then be studied and
characterized\cite{b3,b4}.

To describe the Lyapunov instability for any of such systems, equilibrium
or nonequilibrium, imagine the
deformation of a small phase-space hypersphere $\otimes$, comoving with,
and centered on, a deterministic ``reference trajectory''
$\{ \ q(t),p(t),\zeta(t) \ \}$. As the motion
progresses the hypersphere will deform, at first becoming a rotating
hyperellipsoid with the long-time-averaged exponential growth and decay
rates of the principal axes defining the Lyapunov spectrum
$\{ \ \lambda \ \}$.  Much later the highly-distorted volume element,
through the repeated nonlinear bending and folding caricatured by
Smale's 
``Horseshoe''
mapping, combined with an overall dissipative shrinking, comes to occupy
a multifractal strange attractor.  Though the
apparent dimensionality of this steady-state attractor varies from point
to point, it can be characterized by an overall averaged ``information
dimension'' which is necessarily smaller, for stability of the phase-space
flow (with $\langle \ (d/dt)\ln \otimes \ \rangle < 0$) than is the phase-space
dimension itself\cite{b1,b5}.

The time-averaged exponential growth and decay of phase volume have long
been described by a set of orthonormal Lyapunov vectors $\{ \ \delta \ \}$,
one for each Lyapunov exponent $\lambda$.  Lyapunov {\it spectra}, sets of
time-averaged
local exponents, $\{ \ \lambda = \langle \ \lambda(t) \ \rangle \ \}$,
for a variety of both small and large systems have been determined based
on work pioneered by Stoddard and Ford\cite{b6}, Shimada and
Nagashima\cite{b7}, and Benettin's group\cite{b8}.  The summed-up Lyapunov
spectrum can have thermodynamic significance, corresponding to the loss
rate of Gibbs' entropy (the negative of the entropy gain of the thermal
reservoir regions) in nonequilibrium systems interacting with Nos\'e-Hoover
thermostats, $\sum \lambda = \dot S/k$.  The relative sizes of the
phase-space components of the vector $\delta_1$ associated with the largest
instantaneous Lyapunov exponent, $\lambda_1(t)$, allows instability sources
(``hot spots'', or better, ``regions'') to be located spatially.  This localization of instability
was long studied by Lorenz in his efforts to understand the predictability
of weather.

Recently ideas which had been expressed much earlier by Lorenz\cite{b9,b10,b11}
and Eckmann and Ruelle\cite{b12} have been developed into several
algorithms describing the phase-space deformation with an alternative set of
``covariant vectors'', vectors which ``follow the motion'' in a precisely
time-reversible but somewhat arbitrary way\cite{b13,b14,b15,b16,b17,b18,b19}.
The literature describing
this development is becoming widespread while remaining, for the most part,
overly mathematical (lots of linear matrix algebra) and accordingly hard
to read.  In order better to understand this
work, we apply both the older Gram-Schmidt and the newer covariant
time-reversibility ideas to three simple harmonic-oscillator problems.

To begin, we describe the three example problems in Section II, along with
the usual Gram-Schmidt method for finding local Lyapunov exponents and
corresponding vectors.  Some of the newer covariant approaches are outlined
in Section III.  Numerical results for the three problems, followed by our
conclusions, make up the last two Sections, IV and V.

\section{Lyapunov Spectrum using Lagrange Multipliers}

\begin{figure}
\includegraphics[height=10.5cm,width=6cm,angle=-90]{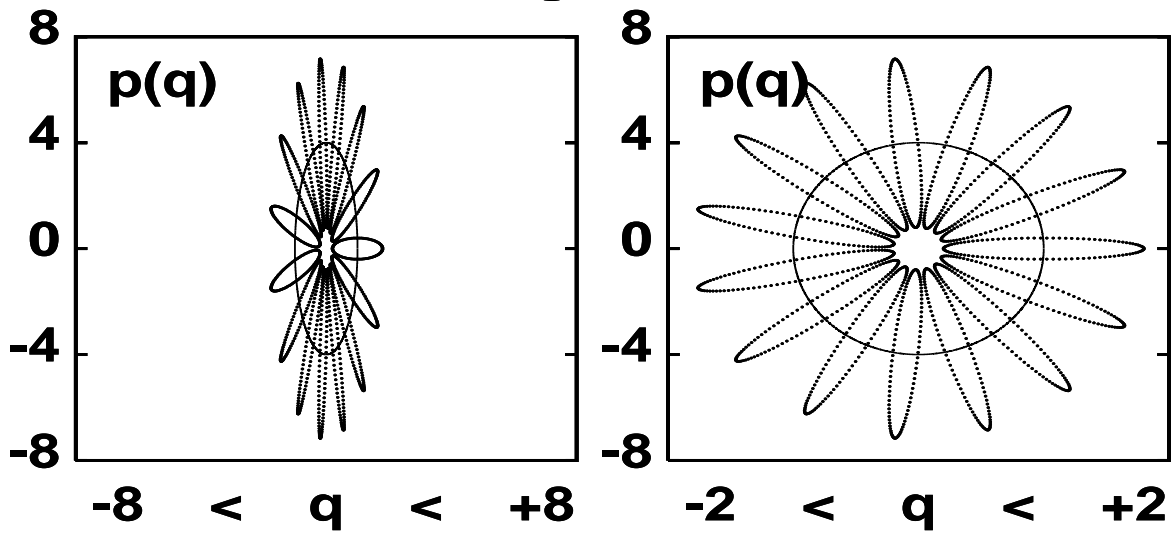}
\vspace{.5in}
\caption{
Two views of a harmonic oscillator orbit with comoving ellipses centered
on the orbit.  The scaled
equations of motion are $\dot q = +ps^{-2} \ ; \ \dot p = -qs^{+2}$. The changing
aspect ratio of the ellipse shown at the left provides nonzero Gram-Schmidt
Lyapunov exponents for the oscillator.  The scaled plot at the right, of
exactly the same data, shows
that the exponents are a consequence of the scale factor $s=2$ discussed
in the text.
}
\end{figure}

\begin{figure}
\includegraphics[height=7cm,width=4cm,angle=-90]{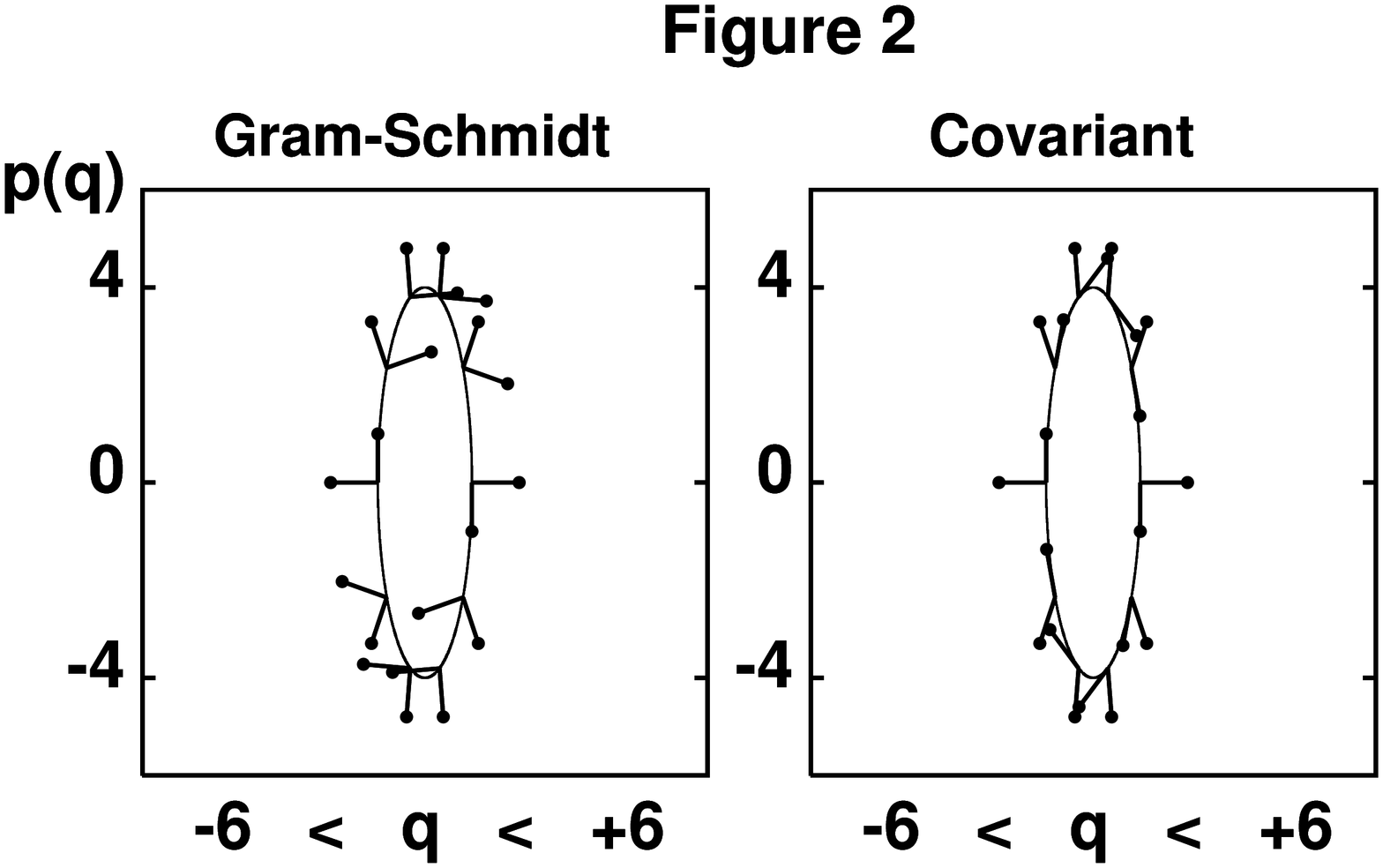}
\vspace{.5in}
\caption{
Periodic phase-space orbit for the harmonic oscillator with scale factor $s=2$.
The Gram-Schmidt vector $\delta_1^f$ is in the radial direction, identical to
the first covariant vector $\delta_1^c$ and necessarily perpendicular
to $\delta_2^f$.  The covariant vector $\delta_2^c$ is parallel to the orbit.  The
Gram-Schmidt vectors are shown here at equally-spaced times and are, for this
special model, identical in the two time directions,
$\{ \ \delta^f \ \} = \{ \ \delta^b \ \}$.
}
\end{figure}

\subsection{Model 1: Simple Harmonic Oscillator}

The models studied here are all harmonic oscillators.  All of them
incorporate variations on the textbook oscillator problem with
coordinate $q$, momentum $p$, and motion equations
$\{ \ \dot q = +p \ ; \ \dot p = -q \ \}$.  The first and simplest
model\cite{b20} can be analyzed analytically.  A $(q,p)$ phase-space
orbit of this oscillator is shown in Figure 1.  Notice that the oscillator
orbit shown there includes an arbitrary scale factor $s$, here chosen
equal to 2.   The corresponding Hamiltonian is
${\cal H}(q,p)$:
$$
2{\cal H} \equiv s^{+2}q^2 + s^{-2}p^2  \ \stackrel{s=2}{\longrightarrow} \
\{ \ \dot q = +(p/4) \ ; \ \dot p = -(4q) \ \} \ \longrightarrow
\ddot q = - q \ .
$$
The ``dynamical matrix'' $D$ for this oscillator describes the evolution of
the set of infinitesimal offset vectors $\{ \ \delta \ \}$ needed to
define the Lyapunov exponents,
$$
\{ \ \delta = (q,p)_{\rm satellite} - (q,p)_{\rm reference} \ \} \ ; \
\{ \ \dot \delta = D \cdot \delta \ \} \ .
$$
In this case the matrix is
$$
D =
\left[
\begin{array}{c  c}
0  &  (1/4) \\
-4  &  0     
\end{array}
\right] \ .
$$
Two local Lyapunov exponents, which reflect the short-term tendency of two
orthonormal  offset
vectors to grow or shrink, can then be defined by the two vector relations:
$$
\dot \delta_1 = D \cdot \delta_1 - \lambda_{11}\delta_1 \ ;
$$
$$
\dot \delta_2 = D \cdot \delta_2 - \lambda_{21}\delta_1  -
\lambda_{22}\delta_2 \ .
$$
We choose the ``infinitesimal'' length of the vectors,
arbitrary in this linear problem, equal to unity for convenience.
Snapshots of the two orthogonal Gram-Schmidt vectors are shown at the
left in Figure 2.

These local exponents have time-averaged values of zero, but sizable
nonzero time-dependent and $s$-dependent fluctuations,
$$
\langle \ (\lambda_{ii} - \langle \ \lambda_{ii} \ \rangle)^2 \ \rangle =
(s^{+1} - s^{-1})^2/2 \ .
$$
The (scalar) Lagrange multipliers $\lambda_{11}$ and $\lambda_{22}$
constrain the {\it lengths} of the offset vectors $\delta_1$ and $\delta_2$
while $\lambda_{21}$ constrains the {\it angle} between the vectors.  The
length-constraining multipliers define the local Gram-Schmidt Lyapunov
exponents.  Their time averages give the global two Lyapunov exponents,
$\lambda_1$ and $\lambda_2$ :
$$
\lambda_1 = \langle \ \delta_1 \cdot D \cdot \delta_1 \ \rangle =
\langle \ \lambda_{11} \ \rangle \ ; \
\lambda_2 = \langle \ \delta_2 \cdot D \cdot \delta_2 \ \rangle =
\langle \ \lambda_{22} \ \rangle \ .
$$

This Lagrange-multiplier approach is the small timestep limit of the
finite-difference Gram-Schmidt approach to orthonormalization\cite{b21,b22}.
For a model with an $N$-dimensional phase space $N$ Lagrange multipliers
$\{ \ \lambda_{ii} \ \}$ constrain the fixed {\it lengths} of the $N$
offset vectors while  $N(N-1)/2$ additional multipliers
$\{ \ \lambda_{ij} \ \}$ are required to keep the $N(N-1)/2$
{\it pairs} of vectors orthogonal.  In the following two subsections we
describe the analytic formulation of the Lagrange-multiplier problem for
examples with three- and four-dimensional phase spaces. For large $N$ the
Gram-Schmidt approach is considerably faster and simpler than the
Lagrange-multiplier one.

\subsection{Model 2: Harmonic Nos\'e-Hoover Oscillator with a
Temperature Gradient $\epsilon$}

The second model, unlike the first, can exhibit long-term chaotic motion,
with three-dimensional phase-space perturbations
$\{ \ \delta q,\delta p,\delta \zeta \ \}$ growing or shrinking exponentially in
time, $\propto e^{\lambda t}$.  Here $\zeta$ is a friction coefficient and
controls the instantaneous changing kinetic temperature $p^2$, so as
to match a specified target temperature $T(q)$\cite{b23,b24}.  Here we
allow the temperature to depend upon the oscillator
coordinate\cite{b25,b26,b27,b28},
$$
1 - \epsilon < T(q) = 1 + \epsilon\tanh(q) < 1 + \epsilon \ .
$$
The temperature gradient makes overall {\it dissipation} possible,
characterized by a shrinking phase-space volume, $\otimes \rightarrow 0$,
and resulting in a strange attractor, with $D_I < 3$, or even a one-dimensional
limit cycle, $D_I=1$.

The $\epsilon$-dependent temperature gradient opens the irresistable
possibility for heat to be
absorbed at higher temperatures than those where it is expelled.  The
Nos\'e-Hoover equations of motion for the nonequilibrium oscillator are :
$$
\{ \ \dot q = p \ ; \ \dot p = - q - \zeta p \ ; \ \dot \zeta = p^2 - T(q) \
\} \ .
$$
The friction coefficient $\zeta$ allows the long-time-averaged kinetic
temperature, proportional to $\langle \ p^2 \ \rangle$, to conform to a
nonequilibrium steady state, characterized by the imposed temperature
profile $T(q)$ :
$$
\langle \ \zeta \ \rangle {\rm \ constant \ }  \longrightarrow
\langle \ p^2 \ \rangle = \langle \ T(q) \ \rangle \ .
$$

Because the motion occurs in a three-dimensional phase space the dynamical
matrix $D$, which governs the motion of phase-space offset vectors
is $3\times 3$ :
$$
D =
\left[
\begin{array}{c  c  c}
(\partial \dot q/\partial q)&(\partial \dot q/\partial p)&
(\partial \dot q/\partial \zeta) \\
(\partial \dot p/\partial q)&(\partial \dot p/\partial p)&
(\partial \dot p/\partial \zeta) \\
(\partial \dot \zeta/\partial q)&(\partial \dot \zeta/\partial p)&
(\partial \dot \zeta/\partial \zeta) 
\end{array}
\right]
=
\left[
\begin{array}{c  c  c}
0  &  1 &  0   \\
-1  &  -\zeta  & -p  \\   
-T^\prime  &  2p &  0   
\end{array}
\right] \ ,
$$
where $T^\prime$ is the derivative of temperature with respect to $q$ :
$$
-(\partial \dot \zeta/\partial q) = T^\prime = \epsilon \cosh^{-2}(q) \ .
$$
For simplicity, we choose these vectors to have unit length.
In addition, {\it six} Lagrange multipliers are required to maintain
the orthonormality of the three offset vectors
$\{ \ \delta_1,\delta_2,\delta_3 \ \}$ :
$$
\left[
\begin{array}{c  c  c }
\lambda_{11} & 0 & 0 \\
\lambda_{21} & \lambda_{22} & 0 \\
\lambda_{31} & \lambda_{32} &  \lambda_{33} 
\end{array}
\right] \ ;
$$

$$
\{ \
\lambda _1 = \langle \ \lambda_{11}(t) \ \rangle \ ; \
\lambda _2 = \langle \ \lambda_{22}(t) \ \rangle \ ; \
\lambda _3 = \langle \ \lambda_{33}(t) \ \rangle \ \} \ .
$$
This model can exhibit chaos or regular behavior, depending on the
initial conditions as well as the maximum temperature gradient
$\epsilon $.  Chaotic solutions for this oscillator are necessarily
numerical, rather than analytical, and are illustrated in
Section III.

\subsection{Model 3: Doubly Thermostated Oscillator with a
Temperature Gradient $\epsilon$}

The Nos\'e-Hoover oscillator, with a single friction coefficient $\zeta$
(Model 2), is not ergodic.  For relatively
small temperature gradients it exhibits an infinity of regular solutions
bathed in a chaotic sea.  For a glimpse of the details see Reference 24.
This geometric three-dimensional complexity can be reduced at the price
of introducing a second thermostat variable.  The resulting
{\it four-dimensional} model, as well as a similar extension
treated in Reference 18, has {\it two} friction coefficients $(\zeta,\xi)$
rather than just one.  With a constant temperature $T$ the resulting ergodic
canonical-ensemble phase-space probability density is
$$
f(q,p,\zeta,\xi) = (1/4\pi^2 T)
e^{-q^2/2T} e^{-p^2/2T} e^{-\zeta^2/2} e^{-\xi^2/2} \ .
$$
The {\it nonequilibrium} multifractal extension of this
model results if (as in Model 2) the temperature depends upon the
oscillator coordinate, $T(q) = 1 + \epsilon \tanh(q)$.  In this
nonequilibrium case there are four equations of motion:
$$
\{ \ \dot q = p \ ; \ \dot p = - q - \zeta p - \xi p^3\ ; \
\dot \zeta = p^2 - T(q) \ ; \ \dot \xi = p^4 - 3p^2T(q) \ \} \ .
$$
The corresponding four-variable dynamical matrix $D$, which controls the
linearized (``tangent-space'') motion, $\dot \delta = D \cdot \delta$, is
$$
D =
\left[
\begin{array}{c  c  c  c}
0  &  1 &  0   &   0  \\
-1  &  [-\zeta - 3\xi p^2]  & -p    &  -p^3   \\   
-T^\prime  &  2p &  0   &  0   \\
-3p^2 T^\prime  &  [4p^3 - 6pT]  & 0     &   0     
\end{array}
\right] \ .
$$

A lower-triangular array of constraining Lagrange Multipliers, can then be
defined.  Just as before, the diagonal multipliers maintain the lengths of the
vectors constant and the off-diagonal multipliers maintain the orthogonality
of the vectors.  For this model, with four offset vectors, we have ten Lagrange
multipliers in all:
$$
\dot \delta_1 = D \cdot \delta_1 - \lambda_{11}\delta_1 \ ;
$$
$$
\dot \delta_2 = D \cdot \delta_2 - \lambda_{21}\delta_1  -
\lambda_{22}\delta_2 \ ;
$$
$$
\dot \delta_3 = D \cdot \delta_3 - \lambda_{31}\delta_1  -
\lambda_{32}\delta_2  - \lambda_{33}\delta_3 \ ;
$$
$$
\dot \delta_4 = D \cdot \delta_4 - \lambda_{41}\delta_1  -
\lambda_{42}\delta_2  - \lambda_{43}\delta_3  - \lambda_{44}\delta_4 \ .
$$

The Gram-Schmidt Lyapunov exponents are the long-time-averaged diagonal
Lagrange Multipliers\cite{b22,b23} :
$$
\lambda_i = \langle \lambda_{ii}(t) \rangle = \langle \delta_{i}
\cdot D \cdot \delta_{i} \rangle / \langle \delta_{i} \cdot \delta_{i}
\rangle \ \equiv \  \langle \delta_{i}
\cdot D \cdot \delta_{i} \rangle \ .
$$
Here, as is usual, we choose the (arbitrary) length of the tangent-space
offset vectors equal to unity:
$\{ \ |\delta| \equiv 1 \ \}$.
The instantaneous off-diagonal elements,
$$
\{ \ \lambda_{ij}(t) = \delta_{i} \cdot D \cdot \delta_{j} +
               \delta_{j} \cdot D \cdot \delta_{i} \ \} \ ,
$$
describe the tendency of the offset vectors to rotate relative to one
another.  This four-dimensional model is already sufficiently complex
to illustrate the distinctions between the Gram-Schmidt and covariant
descriptions of phase-space instabilities. We include numerical results
for this model too, in Section IV.

\section{Covariant Lyapunov Vectors and Their Exponents}

Several loosely-related schemes for evaluating {\it covariant} (as opposed
to Gram-Schmidt) Lyapunov vectors have been described and explored.  Their
proponents are mostly  interested in the mathematics of weather modeling
predictability or in better understanding the statistical mechanical
sensitivity to phase-space perturbations.  Lorenz' famous 1972 talk title:
``Predictability: Does the Flap of a Butterfly's Wings in Brazil
Set off a Tornado in Texas?'' indicates the broad scope of this work\cite{b11}.

Wolfe's 2006 dissertation\cite{b14} and the Kuptsov-Parlitz review\cite{b19}
are particularly useful guides to this work.  Some of the underlying ideas
date back to Lorenz' early studies, in 1965\cite{b9}.  The main idea is
not so different to the old Gram-Schmidt approach and in fact requires
information based on that Gram-Schmidt (or Lagrange multiplier) approach
not only {\it forward} (as is usual), but also {\it backward} (which can
be artificial, violating the Second Law of Thermodynamics) in time.

The main idea is to seek a representative basis set of comoving and
corotating infinitesimal phase-space vectors $ \{ \ \delta^c \ \}$,
($c$ for {\it covariant}) guided by the linearized (treating
$\{ \ q,p,\zeta,\xi \ \}$
as constants in the $D$ matrix) motion equations,
$$
\{ \ \dot \delta^c = D \cdot \delta^c \ \} \ .
$$
The covariant basis vectors follow the linearized flow equations, {\it without}
Lagrange multipliers, and so are generally {\em not} orthogonal.  Provided
that the ``unstable'' manifold,  made up of the phase-space directions
corresponding to longtime expansion, can be usefully distinguished from the
``stable'' manifold, corresponding to directions associated with longtime
contraction, the covariant basis vectors are locally parallel to these two
manifolds.

A ``covariant'' set of vectors would seem {\it not} to require Gram-Schmidt
constraints, or Lagrange multipliers, because orthogonality is not required.
Even so, all the existing computational algorithms which have been developed
to find covariant vectors begin by finding the orthonormal Gram-Schmidt
vectors.  {\em Two sets} of Gram-Schmidt vectors are the next requirement,
the usual forward-in-time vectors $\{ \ \delta^f \ \}$ and the not-so-usual
backward-in-time set $\{ \ \delta^b \ \}$.  This second set of vectors is
obtained by post-processing the time-reversed phase-space trajectory.

In time-reversible energy-conserving Hamiltonian mechanics the
reversed trajectory can be as ``natural'' as the forward trajectory.  In
dissipative systems, with phase-space shrinkage, the stored and reversed
trajectory is typically ``unnatural'', and would violate the Second Law of
Thermodynamics.  Unlike
the Gram-Schmidt forward and backward vectors [ we will denote them by
$\{ \ \delta^f \ \}$ and $\{ \ \delta^b \ \}$ ], the covariant vectors
$\{ \ \delta^c \ \}$ are defined so as to be {\it identical}, at least for
Hamiltonian mechanics, in the two time directions, that is,
``covariant''.  The Gram-Schmidt behavior, with the forward and backward
vectors generally different, is a symmetry breaking whose source is not
at all apparent in the underlying time-symmetric differential equations
of motion. 

The differential equations for the time development of the covariant vectors
account for the simultaneous stretching and rotation of an infinitesimal
phase-space hypersphere.  Evidently the {\it maximum} growth (and maximum
shrinkage) rates correspond to the usual largest Lyapunov exponents forward
and backward in time, $\delta^f_1 = \delta^c_1$ and $\delta^b_1 =
\delta^c_N$ in an $N$-dimensional phase space.  Another covariant offset
vector parallels the local direction of the phase-space velocity, 
$(\dot q,\dot p,\dot \zeta)$.  The remaining $N-3$ covariant vectors require
more work.  The difference between the Gram-Schmidt and
covariant phase-space vectors is illustrated for the simple harmonic
oscillator in Figure 2.

The relatively readable Wolfe-Samelson approach\cite{b16} begins with
the Gram-Schmidt
sets $\{ \ \delta^f,\delta^b \ \}$, forward and backward in time.
Then covariant unit vectors are associated with the long-time-averaged
individual Lyapunov exponents, beginning with the second (or, using time
symmetry, beginning with the next-to-last and working backwards).  The
covariant vectors are then expressed as linear combinations of (some of) the
forward and backward Gram-Schmidt vectors.

The problem is overdetermined, in that $2N$ Gram-Schmidt vectors are used as
bases for the $N$ covariant vectors.  The {\it second} covariant vector can
be written as a linear combination of the first two Gram-Schmidt vectors:
$$
\delta^c_2 = y^f_1\delta^f_1 + y^f_2\delta^f_2 \ .
$$
The constants $\{ \ y^f \ \}$ are then determined by solving a relatively
simple eigenvalue problem.
Likewise, the next-to-last covariant vector can be written in terms of the
first two Gram-Schmidt vectors from the time-reversed trajectory:
$$
\ \delta^c_{N-1} = y^b_1\delta^b_1 + y^b_2\delta^b_2 \ .
$$
In general the $n$th covariant vector can be expressed as a sum of $n$
Gram-Schmidt vectors with the constants $y$ determined by solving a
set of linear equations.  All of these vectors are {\it unit vectors},
with length 1.  The many method variations (using the first few, the
last few, or some of {\it both} sets of Gram-Schmidt vectors) are
discussed in Wolfe's thesis, which is currently available online. The
Appendix of Romero-Bastida, Paz\'o, L\'opez, and Rodriguez' work\cite{b17}, as
well as the Kuptsov-Parlitz review\cite{b19} are also useful guides.

The main difficulty in putting all of this work into perspective
is a result of the fractal/singular nature of the phase-space vectors.
This structure can be traced to bifurcations in the past history and/or in the
future evolution of a particular phase point.  This sensitivity to initial
conditions means that slightly different differential-equation algorithms
can lead to qualitatively different trajectories, making it hard to tell
whether or not two computer programs are consistent with one another.  The best
approach to code validation is to reproduce properties of the covariant
vectors which are insensitive to the integration algorithm.  We will consider
some of these properties for our three example oscillator models.

\section{Results for the Three Harmonic Oscillator Problems}

\subsection{Ordinary One-Dimensional Harmonic Oscillator with Scale
Factor {\it s} = 2}

The first of our three illustrative oscillator problems is an equilibrium
problem in Hamiltonian mechanics, a harmonic oscillator with unit frequency
and with a scale factor $s$ set equal to 2 :
$$
\ddot q = -q \longleftarrow
2{\cal H} \equiv s^{+2}q^2 + s^{-2}p^2  \ \stackrel{s=2}{\longrightarrow} \
\{ \ \dot q = +(p/4) \ ; \ \dot p = -(4q) \ \}
\longrightarrow \ddot q = - q \ .
$$
The local Lyapunov exponents for such an oscillator depend upon $s$ and
vanish in the usual case where $s=1$ .

A $(q,p)$ phase-space plot of a typical orbit for $s=2$ was shown in Figure
1.  The oscillator orbit we choose to analyze is the ellipse shown there:
$$
\{ \ q = \cos(t) \ ; \ p = -4\sin(t) \ \} \ ;
\ 16q^2 + p^2 = 16 = 2{\cal H} \ .
$$
The local Lyapunov exponents describe the instantaneous growth rates of
infinitesimal ``offset vectors'' comoving with the flow.  These are of
three kinds, orthonormal vectors moving forward in time $\{ \ \delta^f \ \}$,
orthonormal vectors moving backward in time $\{ \ \delta^b \ \}$, and special
covariant vectors $\{ \ \delta^c \ \}$
whose forward and backward orientations in phase space are identical. The
initial direction of each of these vectors is arbitrary because a reversed
(antiparallel) vector $(+\delta \longleftrightarrow -\delta)$
satisfies exactly the same equations and definitions.

For the simple scaled harmonic oscillator it is possible to solve analytically
for the
Gram-Schmidt vectors forward and backward in time as well as the covariant
vectors and all the corresponding local exponents,
$$
\{ \ \delta^f(t),\delta^b(t),\delta^c(t) \ \} \longrightarrow
\{ \ \lambda^f(t),\lambda^b(t),\lambda^c(t) \ \} \ ,
$$
once the initial conditions are specified.  See Figure 2 for an illustration
of the following choice of initial values, at $t=0$ :
$$
(q,p) = (1,0) \ ; \ \delta^f_1 = \delta^b_1 = \delta^c_1 = (1,0) \ ; \
\delta^f_2 = \delta^b_2 = \delta^c_2 = (0,1) \ .
$$
In the course of the motion, with period $2\pi$, the first forward
Gram-Schmidt vector $\delta^f_1$, which is generally identical to the
first covariant vector, turns out to be {\it always} radial.  This vector
is necessarily perpendicular to $\delta^f_2(t)$.  The vectors $\delta^f_2$
and $\delta^c_2$ generally differ.  The {\it second } covariant vector is
a unit vector which remains always parallel or antiparallel (and hence
``covariant'') to the orbit,
$$
\delta^c_2 = \pm(\dot q,\dot p)/|(\dot q,\dot p)| \ .
$$
Generally, all the covariant vectors satisfy the {\it linearized}
small-$\delta $ motion equations.  Because the oscillator motion equations
{\it are} linear, these oscillator results are correct for finite, not
just infinitesimal, vectors.

The Gram-Schmidt Lyapunov exponents can be calculated as Lagrange multipliers
which maintain the orthonormal relations between
$\{ \ \delta_1(t),\delta_2(t) \ \}$ in both the forward and the backward
directions. For this simple problem the
vectors are unchanged by time reversal, $\delta^f_1 \equiv \delta ^b_1 \ ; \
\delta^f_2 \equiv \delta ^b_2$.  The first vector is not constrained in
orientation, but is restricted to constant length by the Lagrange multiplier
$\lambda_{11}$:
$$
\{ \ \dot \delta q_1 = (1/4)\delta p_1 - \lambda_{11}\delta q_1 \ ; \
\dot \delta p_1 = -4\delta q_1 - \lambda_{11}\delta p_1 \ \} \ \longrightarrow
\ \lambda_{11} = - (15/4)\delta q_1\delta p_1.
$$
Using $\lambda_{11}$ in either the $\dot \delta q_1$ or  $\dot \delta p_1$
equation gives a differential
equation for the unit vector $\delta _1 = (\delta q_1,\delta p_1)$, which
simplifies with the definition of a new variable, the angle $\theta$:
$$
\dot \delta q_1 = (1/4)\delta p_1 + (15/4)\delta p_1 \delta q_1^2 \ ; \
\dot \delta p_1 = -4\delta q_1 + (15/4)\delta q_1 \delta p_1^2
$$
$$
[ \ \delta q_1 \equiv \cos(\theta) \ ; \ \delta p_1 = -\sin(\theta) \ ]
 \longrightarrow
$$
$$
\dot \theta = (1/4) + (15/4)\cos^2(\theta) = 4 - (15/4)\sin^2(\theta)
\longrightarrow \theta = \arctan (4\tan(t)) \ .
$$
Although the probability density for $\delta q$ [ or $\delta p$ ] diverges, 
whenever $\dot q = 0$ [ or whenever $\dot p = 0$ ], the probability density
for $\theta $, which describes the nonuniform phase-space rotary motion,
is well-behaved, as shown in Figure 3.

The instantaneous Lyapunov vectors, both Gram-Schmidt and covariant, from
the orbit shown in Figure 2, are shown as functions of time in Figure 4.
The exponents  are:
$$
\lambda^f(t)  = \lambda^b(t) = \pm (-15/4)\sin(\theta)\cos(\theta) \ ;
\ \theta = \arctan (4\tan (t)) \ ,
$$
with the second covariant exponent given by the strain rate parallel to the
orbit:
$$
d\ln v_\parallel/dt = -15\sin(t)\cos(t)/[\sin^2(t) + 16\cos^2(t)] \ .
$$
Figure 4 shows that this covariant exponent is $90^o$ out of phase with
the exponent associated with $\delta^f_1(t)$.

\begin{figure}
\includegraphics[height=7cm,width=4cm,angle=-90]{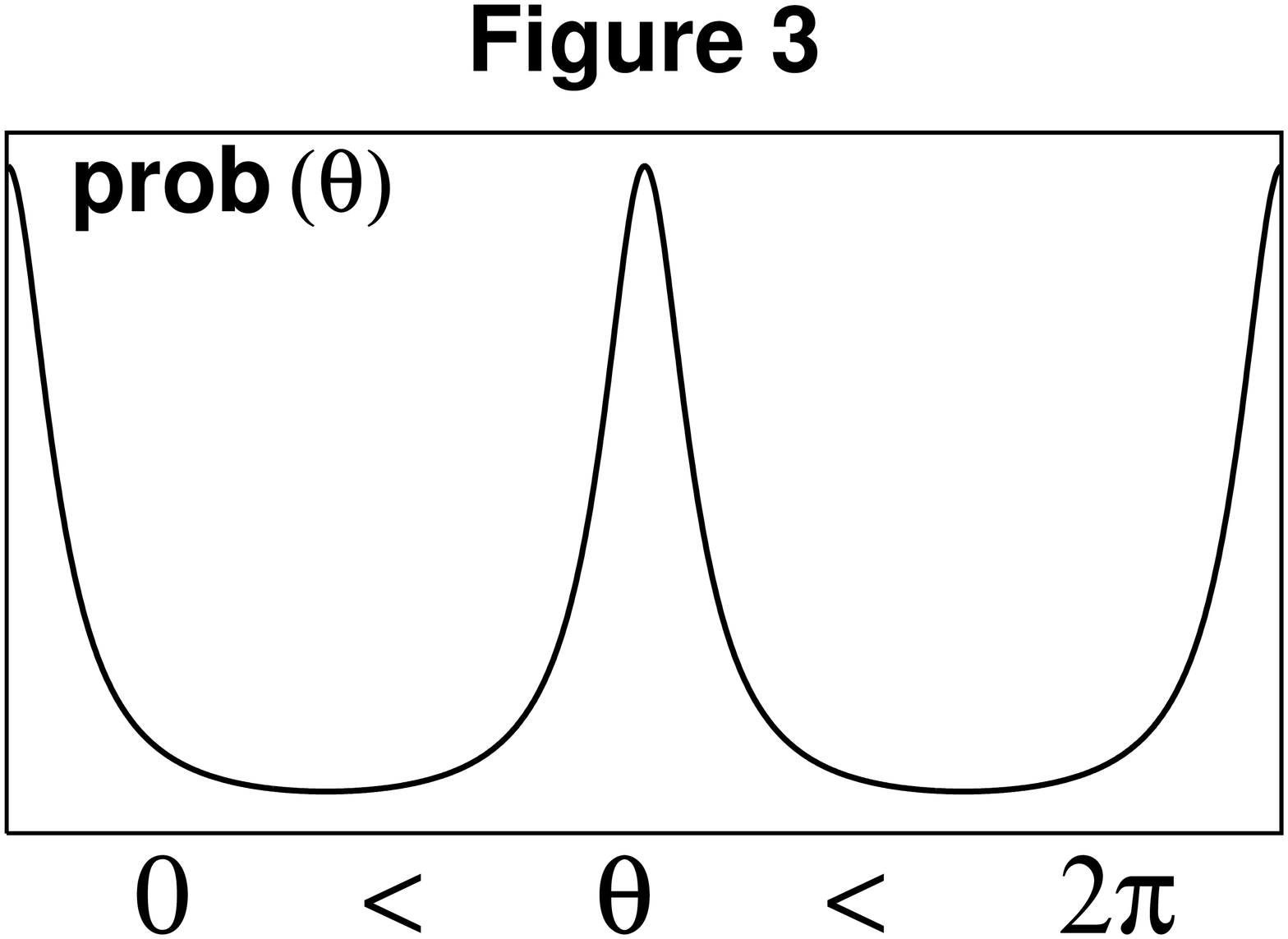}
\vspace{.5in}
\caption{
Probability density for the angle $\theta = \arctan(4\tan (t))$ for the
scaled harmonic oscillator.
}
\end{figure}

\begin{figure}
\includegraphics[height=7cm,width=4cm,angle=-90]{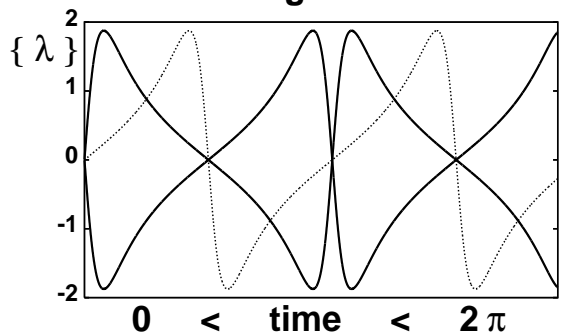}
\vspace{.5in}
\caption{
Lyapunov exponents for the harmonic oscillator with scale factor 2.  The
Gram-Schmidt exponents are the two heavy lines, and sum to zero.  The
{\it second} covariant exponent, $\lambda^c_{22}(t) = (d\ln v_\parallel/dt)$
along the orbit, is the lighter dashed line.
}
\end{figure}

Because the oscillator is not chaotic, these detailed results depend upon
the choice of initial conditions.
The {\it general} features illustrated by this problem include (1) the
orthogonality of the Gram-Schmidt vectors, (2) the possibility of obtaining
a ``reversed'' trajectory by using a stored forward trajectory, changing only
the sign of the timestep $+dt \rightarrow -dt$, (3) the identity of the first
Gram-Schmidt vector with the first covariant vector, and (4) the identity of
the trajectory direction with the covariant vector corresponding to the
time-averaged exponent $\langle \ \lambda_\parallel \ \rangle \equiv 0$. In the
next problem we study the directions of the forward and backward vectors 
and show that they typically differ, though the {\it time-averaged} exponents
(for a sufficiently long calculation) agree, apart from their signs.

\subsection{Three-Dimensional $(q,p,\zeta)$ Nos\'e-Hoover Oscillator }

\begin{figure}
\includegraphics[height=7cm,width=4cm,angle=-90]{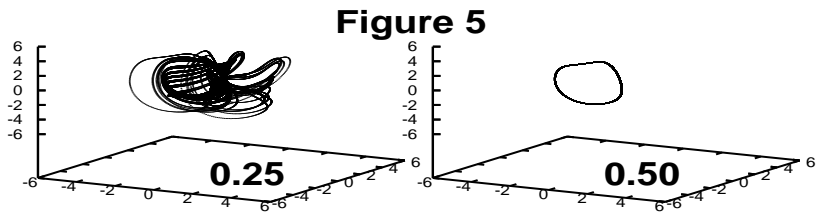}
\vspace{.5in}
\caption{
Phase-space trajectories for a Nos\'e-Hoover heat conducting oscillator
with the coordinate-dependent temperature, $T = 1 + \epsilon\tanh(q)$.
The strange attractor at the left corresponds to a maximum temperature
gradient $\epsilon = 0.25$ while the limit cycle at the right
corresponds to $\epsilon = 0.50$.  Here the coordinate $q$ increases
from left to right, the momentum $p$ from front to back, and the friction
coefficient $\zeta$ from bottom to top.
}
\end{figure}

Here we consider the second model oscillator.  Its {\it nonequilibrium}
temperature profile,
$$
T(q) = 1 + \epsilon \tanh(q) \ ,
$$
provides both chaotic and regular solutions, depending on the strength
of the temperature gradient.  Trajectory segments for two values of
$\epsilon$ (the maximum value of the temperature gradient) are shown
in Figure 5. For the smaller value $\epsilon = 0.25$, the {\it second}
Lyapunov exponent vanishes, $\lambda^f_2 = \langle \ \lambda^f_{22}(t)
\  \rangle = 0$.  Its probability density, shown in Figure 6, is quite
different to that of the second covariant exponent, $\lambda^c_{22}(t)$, which
remains parallel to the trajectory direction at all times.
For the larger limit-cycle value, $\epsilon = 0.50$, the longtime
orbit is a limit cycle, with period 8.650.  On this limit cycle the largest
Lyapunov exponent is identical to the largest covariant exponent, and
necessarily vanishes.  The time dependence of this exponent,
$\lambda^f_{11}(t) = \lambda^c_{11}(t)$, with
$$
\langle \ \lambda^f_{11}(t) \ \rangle =
\langle \ \lambda^c_{11}(t) \ \rangle = \lambda^f_1 = \lambda^c_1 = 0 \ ,
$$
is shown in Figure 7.

\begin{figure}
\includegraphics[height=7cm,width=4cm,angle=-90]{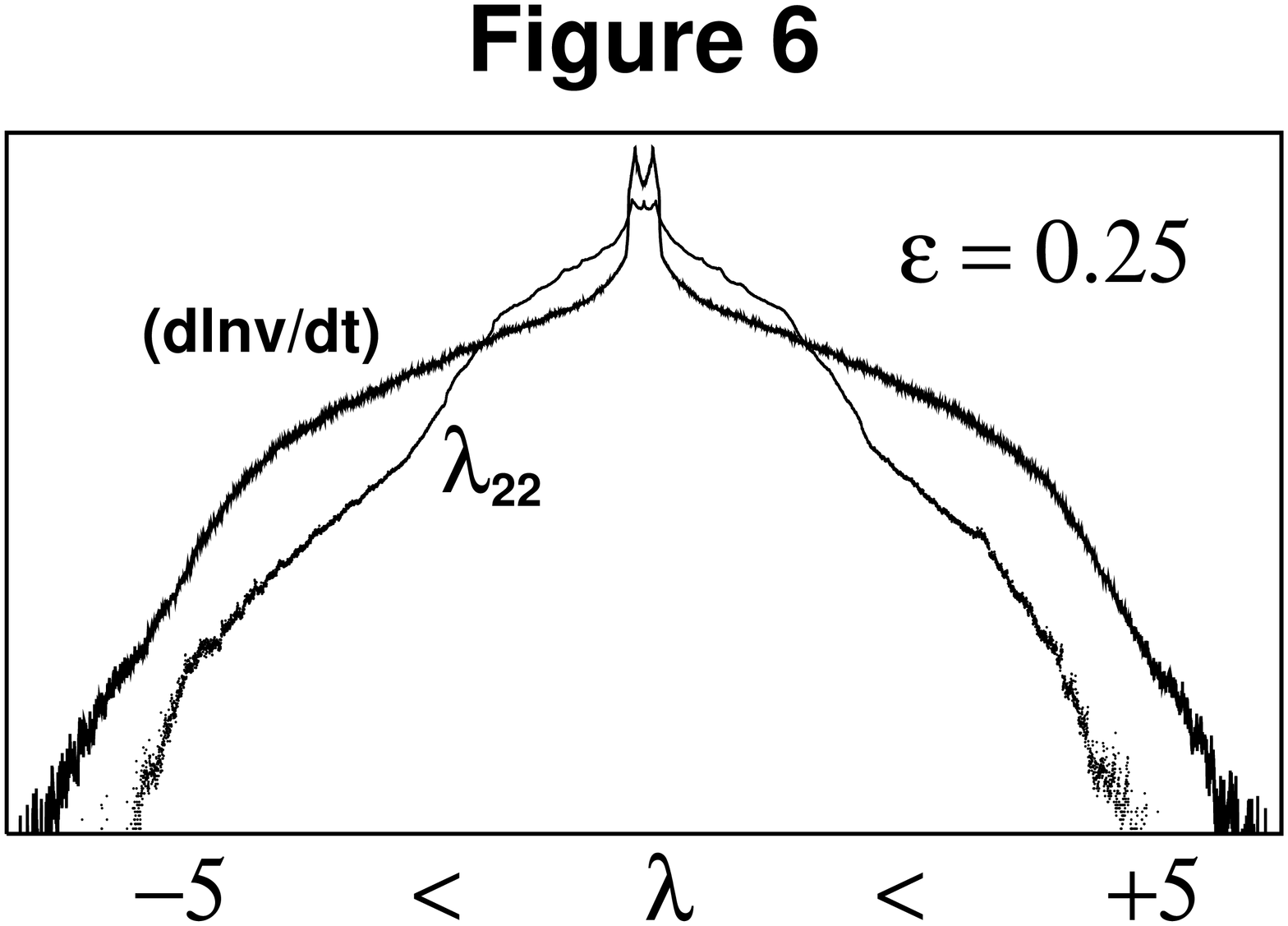}
\vspace{.5in}
\caption{
Logarithms of probability densities, over about six decades, for the orbital
strain rate, $\lambda^c_{22}(t) = (d\ln v/dt)$ and the corresponding instantaneous
Gram-Schmidt Lyapunov exponent, $\lambda^f_{22}$  for the chaotic oscillator
with $\epsilon = 0.25$.
}
\end{figure}

\begin{figure}
\includegraphics[height=7cm,width=4cm,angle=-90]{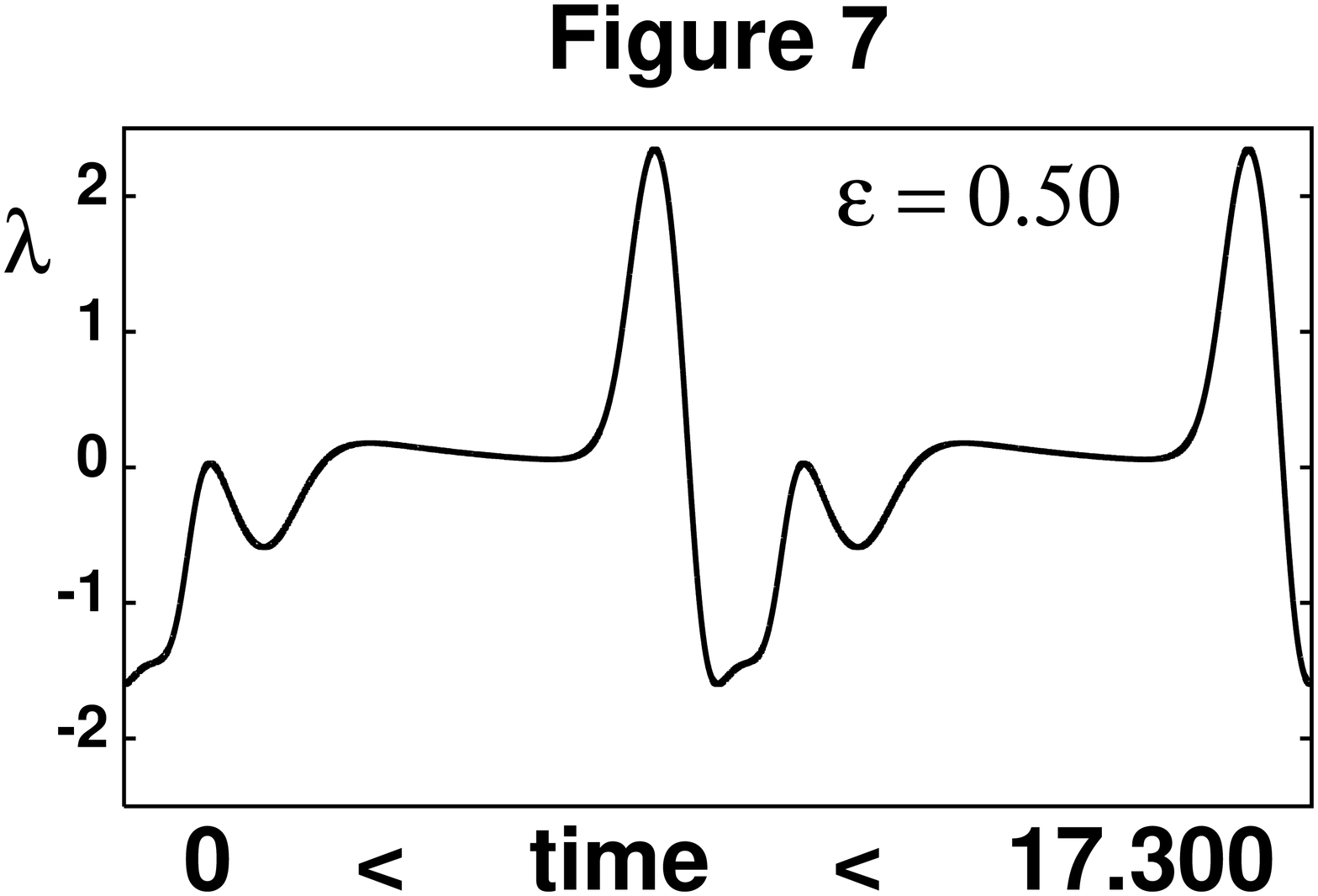}
\vspace{.5in}
\caption{
Instantaneous value of the orbital strain rate for the limit cycle with
$\epsilon = 0.50$.  The strain rate is identical to the corresponding
instantaneous Gram-Schmidt Lyapunov exponent, $\lambda^f_{11}$.
}
\end{figure}

The three-dimensional system of motion equations,
$$
\{ \ \dot q = p \ ; \ \dot p = -q -\zeta p ;
\ \dot \zeta = p^2 - T(q) \ \} \ ,
$$
can be ``reversed'' in either of two ways: (1) replace $+dt$ by $-dt$
in the fourth-order  Runge-Kutta integrator or (2) leaving $dt > 0$
unchanged, change the {\it signs} of the momentum $p$ and the friction
coefficient $\zeta$ so that the underlying physical system traces its
coordinate history backward in time, $q(+t) \rightarrow q(-t)$.  In either
case longtime stability with decreasing time requires that the dissipative
forward trajectory be stored and reused.  If the trajectory is {\it not}
stored then the reversed motion abruptly leaves the unstable reversed orbit.
Figure 8 shows the jump from the illegal reversed motion, violating the
Second Law of Thermodynamics, to a stable obedient motion after sixteen
circuits of the illegal cycle.

\begin{figure}
\includegraphics[height=7cm,width=6cm,angle=-90]{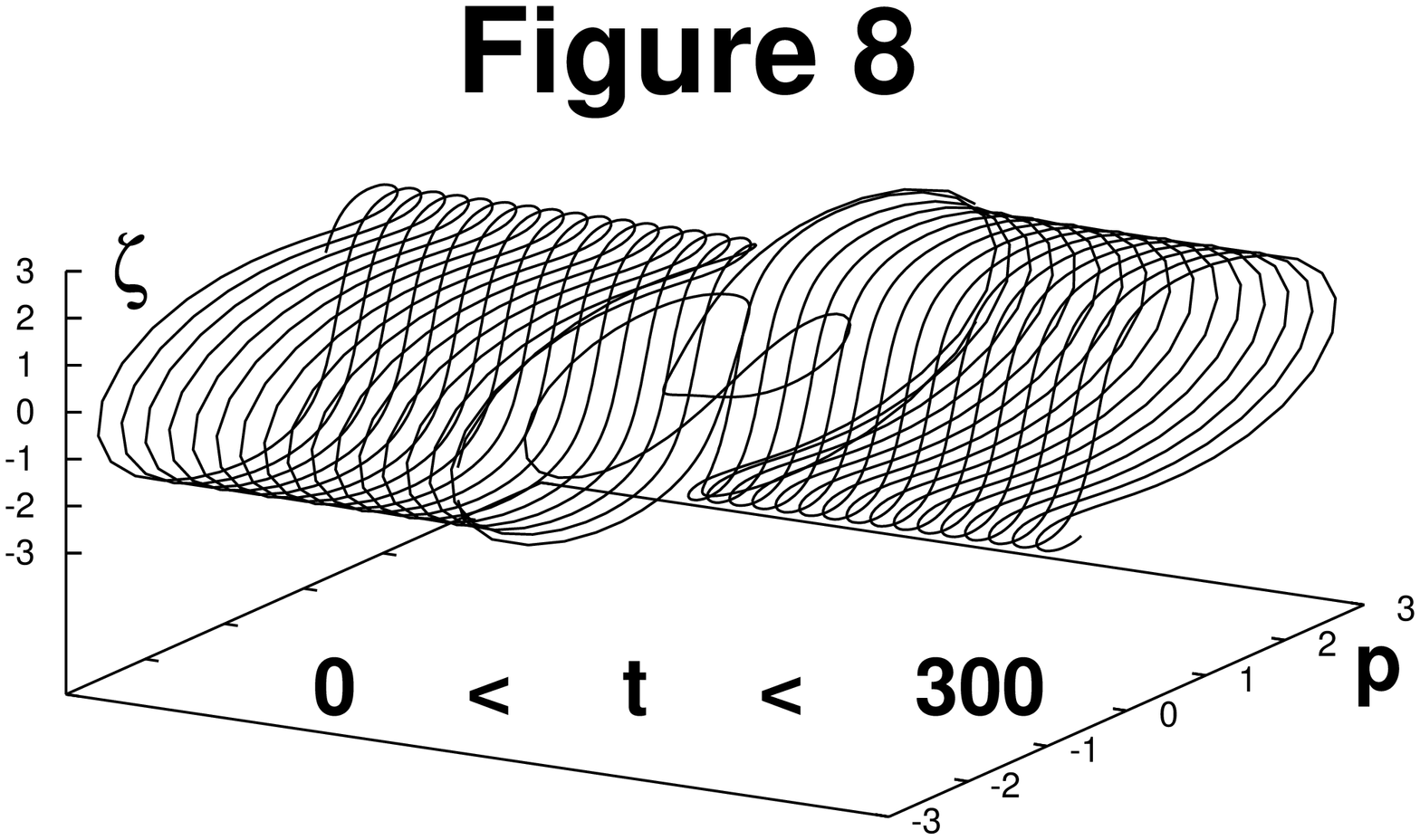}
\vspace{.5in}
\caption{
Time history of a Nos\'e-Hoover oscillator in a temperature gradient,
$\epsilon = 0.50$.  The variation of $\{ \ p,\zeta \ \}$ in time just
after the reversal at time zero is shown.  After 16 ``wrongway'' limit
cycles, during which entropy is absorbed by the oscillator, the
thermodynamically unstable trajectory appears to jump to a cycle which
obeys the Second Law of Thermodynamics.  The jump time of 150 closely
corresponds to the estimate appropriate to 48-bit precision, a cycle
period of 9, and the known Lyapunov exponent of 0.222:
$e^{\lambda t} = e^{0.222t} = 2^{48} \longrightarrow t = 150 \simeq
16 \times 9$.
}
\end{figure}

For simplicity of notation and description we have adopted the first
reversibility definition  throughout the present work.  It is necessary
to recognize that the ``motion'' (forward in time) we analyze {\it is}
dissipative, though time-reversible, while the reversed motion (backward
in time and contrary to the Second Law of Thermodynamics) is actually so
unstable ( because $\langle \ \dot \otimes \ \rangle > 0$ ) as to be
unobservable unless the trajectory has been stored in advance.

The instability of the reversed trajectory, leading to symmetry breaking, can
be understood by considering the flow of probability density in phase space.
The comoving probability density $f(q,p,\zeta,t)$ obeys the analog of Liouville's
continuity equation.  At any instant of time the changing phase-space
probability density responds to the friction coefficient $\zeta$ :
$$
[\dot f/f] \equiv (\partial \dot q/\partial q) +(\partial \dot p/\partial p)
+ (\partial \dot \zeta /\partial \zeta) = 0 + \zeta + 0 =
-\dot \otimes/\otimes =
-\lambda^f_{11}(t) - \lambda^f_{22}(t) - \lambda^f_{33}(t)  \ .
$$
Phase-volume expands/contracts when $\zeta$ is negative/positive.  The
{\it time-averaged} value of the friction coefficient $\zeta$ is necessarily
positive, and reflects the heat transfer (from larger to smaller values
of the oscillator coordinate $q$) consistent with the Second Law of
Thermodynamics.

Because this Nos\'e-Hoover problem is chaotic, no analytic solution is
available, making it necessary to compute the vectors and exponents numerically.
Lyapunov exponents have long been determined numerically.  Spotswood
Stoddard and Joseph Ford's pioneering implementation was generalized
by Shimada, Nagashima, and by Bennetin's group.  These latter authors used
Gram-Schmidt rescaling of phase-space offset vectors.  This continuous
limit of this numerical approach was formalized by introducing Lagrange
multipliers $\{ \ \lambda_{ij} \ \}$ to constrain
the offset vectors' orthonormality\cite{b21,b22}.

This problem raises a computational question: How best to distinguish the
usual Gram-Schmidt vectors $\{ \ \delta^f \ \}$ from the covariant ones
$\{ \ \delta^c \ \}$?  Some quantitative criterion is required, at the
least for reproducibility if not for clarity.  In the case
of the thermostated oscillator we have computed the mean-squared projections
into $(q,p,\zeta)$ space for the three Gram-Schmidt basis vectors, both
forward and backward in time.  The results are essentially the same for
either time direction:
$$
(0.22,0.29,0.49),(0.27,0.38,0.35),(0.51,0.32,0.17) {\rm \ for \ } \epsilon =
0.25 \ ;
$$
$$
(0.22,0.25,0.53),(0.27,0.41,0.32),(0.51,0.34,0.15) {\rm \ for \ } \epsilon =
0.50 \ .
$$
Evidently, at least for this problem, there is little geometric dependence
of the Gram-Schmidt vectors on the presence or absence of chaos.  On the
other hand, the second
covariant vector (parallel to the flow) has mean-squared components
$(0.21,0.28,0.51)$ for $\epsilon = 0.25$ and $(0.22,0.25,0.53)$ for
$\epsilon = 0.50$, and so is significantly different to the Gram-Schmidt
$\delta^f_2$ or $\delta^b_2$. A shortcoming of work on the covariant vectors
so far has been the lack of quantitative information concerning them.  This
is an undesirable situation, as it makes checking computations problematic.

We turn next to a four-dimensional problem.  Although the first and
last Gram-Schmidt vectors and the trajectory direction give three of the
four covariant vectors, the fourth requires new ideas, and illustrates the
general case of determining a complete set of covariant basis vectors.

\subsection{Four-Dimensional $(q,p,\zeta,\xi)$ Doubly-Thermostated Oscillator }

Problems with four or more phase-space dimensions require the mathematics,
mostly linear algebra, of covariant vectors for their solution.  Here we
choose a doubly-thermostated harmonic oscillator (two thermostat variables,
$\zeta$ and $\xi$), again with a nonequilibrium temperature profile,
$T = 1 + \tanh(q)$.  The equations of motion (which give the canonical
phase-space distribution characteristic of the temperature $T$ when
$\epsilon$ vanishes) are:
$$
\{ \ \dot q = p \ ; \ \dot p = -q - \zeta p - \xi p^3 \ ; \ 
\dot \zeta = p^2 - T \ ; \ \dot \xi  = p^4 - 3p^2T \ \} \ .
$$
Again $T$ is the kinetic temperature, the time-averaged value of $p^2$.
$\zeta$
and $\xi$ are time-reversible thermostat variables which control the
second and fourth moments of the velocity distribution.  At equilibrium
the solution of these motion equations is the complete canonical phase-space
distribution.  For $T=1$ this stationary distribution has the form:
$$
f(q,p,\zeta,\xi) \propto \exp [-(q^2 + p^2 + \zeta^2 + \xi^2)/2] \ .
$$
This oscillator system is ergodic.  When the temperature $T$ is made to
depend upon the coordinate $q$, a nonequilibrium strange attractor results.
We can give some detailed results for the well-studied\cite{b25,b26,b27}
special case $0 < T(q) \equiv 1 + \tanh(q) <2$.  For this nonequilibrium
problem the Lyapunov spectrum forward in time is already known,
$$
\{ \ \lambda^f \ \} = \{ \ +0.073, 0.000, - 0.091, - 0.411 \ \} \longrightarrow
\{ \ \lambda^b \ \} = \{ \ +0.411,+0.091, 0.000, - 0.073 \ \} \ ,
$$
as is also the strange attractor's phase-space information dimension,
$D_I$ = 2.56, reduced by 1.44 from the equilibrium case ($\epsilon = 0$),
where the spectrum is symmetric:
$$
\{ \lambda \} = \{ \ +0.066, 0.000, 0.000, - 0.066 \ \} \ .
$$
This information dimension result is in definite violation of the
Kaplan-Yorke conjecture\cite{b27},
$$
D_I = 2.56 \stackrel{?}{\equiv} D_{\rm KY} = 2 + (0.073/0.091) = 2.80 \ .
$$

The Gram-Schmidt vectors $\{\ \delta^f,\delta^b \ \}$ for this problem
are easily obtained using the Lagrange Multiplier approach, and give three
of the four covariant vectors:
$$
\delta^c_1 = \delta^f_1 \ ;
\ \delta^c_2 = v/|v| \ ;
\ \delta^c_4 = \delta^b_1 \ ,
$$
where $v$ is the phase-space trajectory velocity,
$(\dot q, \dot p, \dot \zeta, \dot \xi )$.

Wolfe and Samelson show that the only missing covariant vector, $\delta^c_3$,
can be expressed as a linear combination of the first three forward (or the
last two backward) $\delta$ vectors.
Either choice requires solving an eigenvalue problem numerically
and both choices lead to exactly the same covariant vector $\delta^c_3$,
even if the Gram-Schmidt vectors are not perfectly converged.

A comparison of the various local Lyapunov exponents is detailed in
Figures 9-12.  We have ordered the {\it thickness} of the plotted lines
throughout, in the order $1>2>3>4$ so that the thickest line represents the
distribution for $\lambda^f_{11}(t)$ (mean value +0.073) and, in the
Backward plot,  $\lambda^b_{11}(t)$ (mean value +0.411).  Visually, the
data for two billion points, shown here, are indistinguishable from data
for two hundred million points, shown in Version 1 of this work.

With a single exception, the vanishing Gram-Schmidt exponent derived from
$\delta^f_{22}(t)$, the data suggest a fractal nature
for the probability densities of the various exponents.  Figure 9
compares the probability distributions for $\lambda^f_{22}(t)$ and the
covariant exponent $\lambda^c_{22}(t)$ corresponding to motion along the
trajectory.  The Gram-Schmidt histogram is {\it much smoother} than that
of its covariant cousin.  At the moment we have no explanation for
this interesting qualitative difference.  

Otherwise the local exponent distributions are qualitatively similar---
large fluctuations compared to the actual exponent values.  The extra work
associated with the local covariant spectrum cannot be justified based on
these data.  It should be noted that the Gram-Schmidt exponents forward and
backward in time are quite different, reflecting the difference between the
future and the past trajectories.

\begin{figure}
\includegraphics[height=7cm,width=4cm,angle=-90]{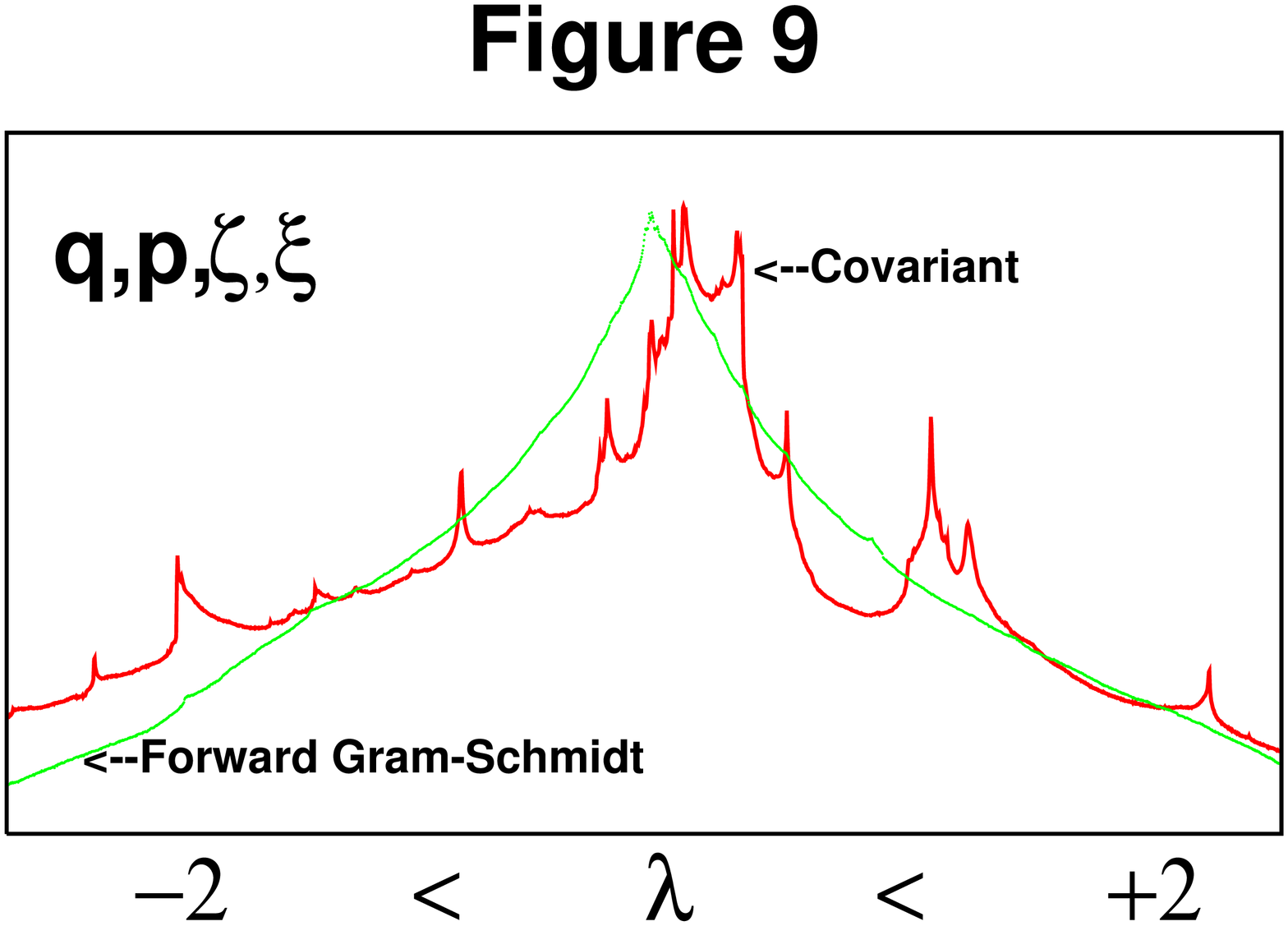}
\vspace{.5in}
\caption{
Probability densities for $\lambda^f_{22}(t)$ and $\lambda^c_{22}(t)$.  The
logarithmic ordinate scale covers two decades with data from the central
two billion out of four billion fourth-order Runge-Kutta timesteps of 0.001
each.
}
\end{figure}

To make it possible for the reader to coordinate our results with those of
Wolfe and Samelson\cite{b16}, we write a set of equations which can be solved in
order to find the third covariant vector $\delta^c_3$ as an expansion in
the {\it forward-in-time} Gram-Schmidt vectors:
$$
M^f \cdot y = 0 \ ; \
M^f =
\left[
\begin{array}{c  c  c}
\sum(\delta^f_1\cdot\delta^b_k)(\delta^b_k\cdot\delta^f_1) \ &
\sum(\delta^f_1\cdot\delta^b_k)(\delta^b_k\cdot\delta^f_2) \ &
\sum(\delta^f_1\cdot\delta^b_k)(\delta^b_k\cdot\delta^f_3)   \\  
 
\sum(\delta^f_2\cdot\delta^b_k)(\delta^b_k\cdot\delta^f_1) \ &
\sum(\delta^f_2\cdot\delta^b_k)(\delta^b_k\cdot\delta^f_2) \ &
\sum(\delta^f_2\cdot\delta^b_k)(\delta^b_k\cdot\delta^f_3)   \\ 
  
\sum(\delta^f_3\cdot\delta^b_k)(\delta^b_k\cdot\delta^f_1) \ &
\sum(\delta^f_3\cdot\delta^b_k)(\delta^b_k\cdot\delta^f_2) \ &
\sum(\delta^f_3\cdot\delta^b_k)(\delta^b_k\cdot\delta^f_3)   
\end{array}
\right] \ .
$$
Here the sums include the two values of $k=1,2$.  The numbering system,
though arbitrary, is crucial.  Forward in time the
correspondence of the vectors with the long-time-averaged Lyapunov
exponents is
$$
\delta^f_1 \rightarrow + 0.073 \ ; \ \delta^f_2 \rightarrow 0.000\ ; \
\delta^f_3 \rightarrow -0.091 \ ; \ \delta^f_4 \rightarrow -0.411 \ .
$$
{\it Backward} in time the ordering is the same with the {\it signs changed}:
$$
\delta^b_1 \rightarrow -0.073 \ ; \ \delta^b_2 \rightarrow  0.000\ ; \
\delta^b_3 \rightarrow +0.091 \ ; \ \delta^b_4 \rightarrow +0.411 \ .
$$

Of course the {\it third} covariant vector going forward could equally
well be viewed as the second going backward using an expansion in terms 
of the {\it backward-in-time} Gram-Schmidt vectors.  This alternative
approach results in a smaller matrix {\it without} the sum over $k$:
$$
M^b \cdot y = 0 \ ; \
M^b =
\left[
\begin{array}{c  c}
(\delta^b_1\cdot\delta^f_1)(\delta^f_1\cdot\delta^b_1) \ &
(\delta^b_1\cdot\delta^f_1)(\delta^f_1\cdot\delta^b_2) \ \\  
 
(\delta^b_2\cdot\delta^f_1)(\delta^f_1\cdot\delta^b_1) \ &
(\delta^b_2\cdot\delta^f_1)(\delta^f_1\cdot\delta^b_2) \ 
\end{array}
\right] \ .
$$
Here the ordering of the Gram-Schmidt vectors follows this same new
ordering of the backward vectors:
$$
\delta^b_1 \rightarrow +0.411 \ ; \ \delta^b_2 \rightarrow +0.091 \ ; \
\delta^b_3 \rightarrow 0.000 \ ; \ \delta^b_4 \rightarrow -0.073 \ .
$$
The vectors forward in time follow the {\it same} ordering with the
{\it signs changed}.

Once the ordering of the vectors is correctly negotiated one can find the
eight Gram-Schmidt vectors (four in each time direction) and the four
covariant vectors (the same in either time direction).  For comparison
we show the probability densities for the vectors in Figures 10, 11, and
12.

\begin{figure}
\includegraphics[height=7cm,width=4cm,angle=-90]{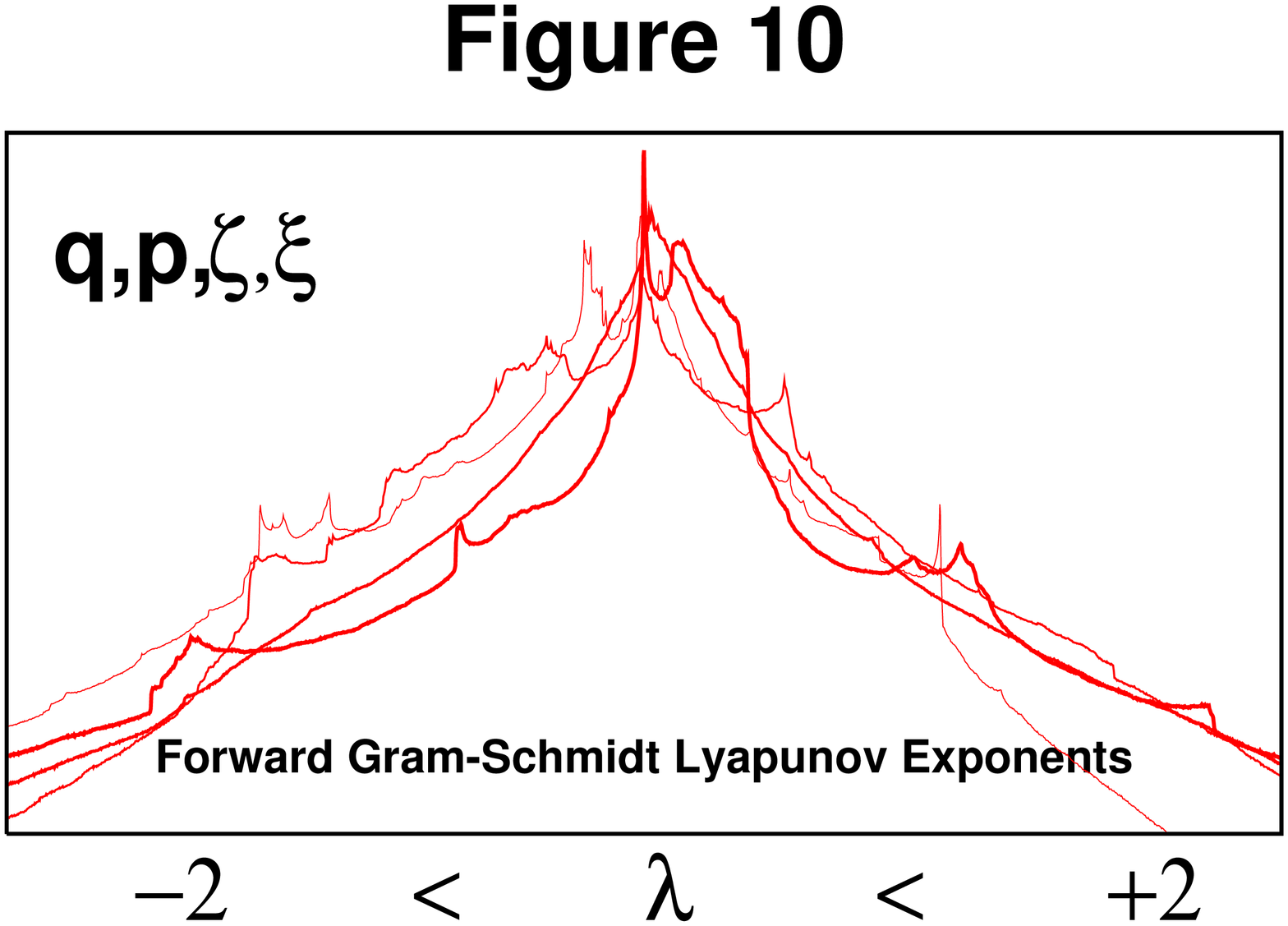}
\vspace{.5in}
\caption{
Probability density for the four forward-in-time Lyapunov exponents,
$\{ \ \lambda^f_{ii}(t) \ \}$.  Notice that the distribution for the
largest of the exponents (widest line) is the same as that for the
covariant Lyapunov exponent $\lambda^c_{11}(t)$.  The logarithmic
ordinate scale covers two decades with data from the central
two billion out of four billion fourth-order Runge-Kutta timesteps of 0.001
each.
}
\end{figure}

\begin{figure}
\includegraphics[height=7cm,width=4cm,angle=-90]{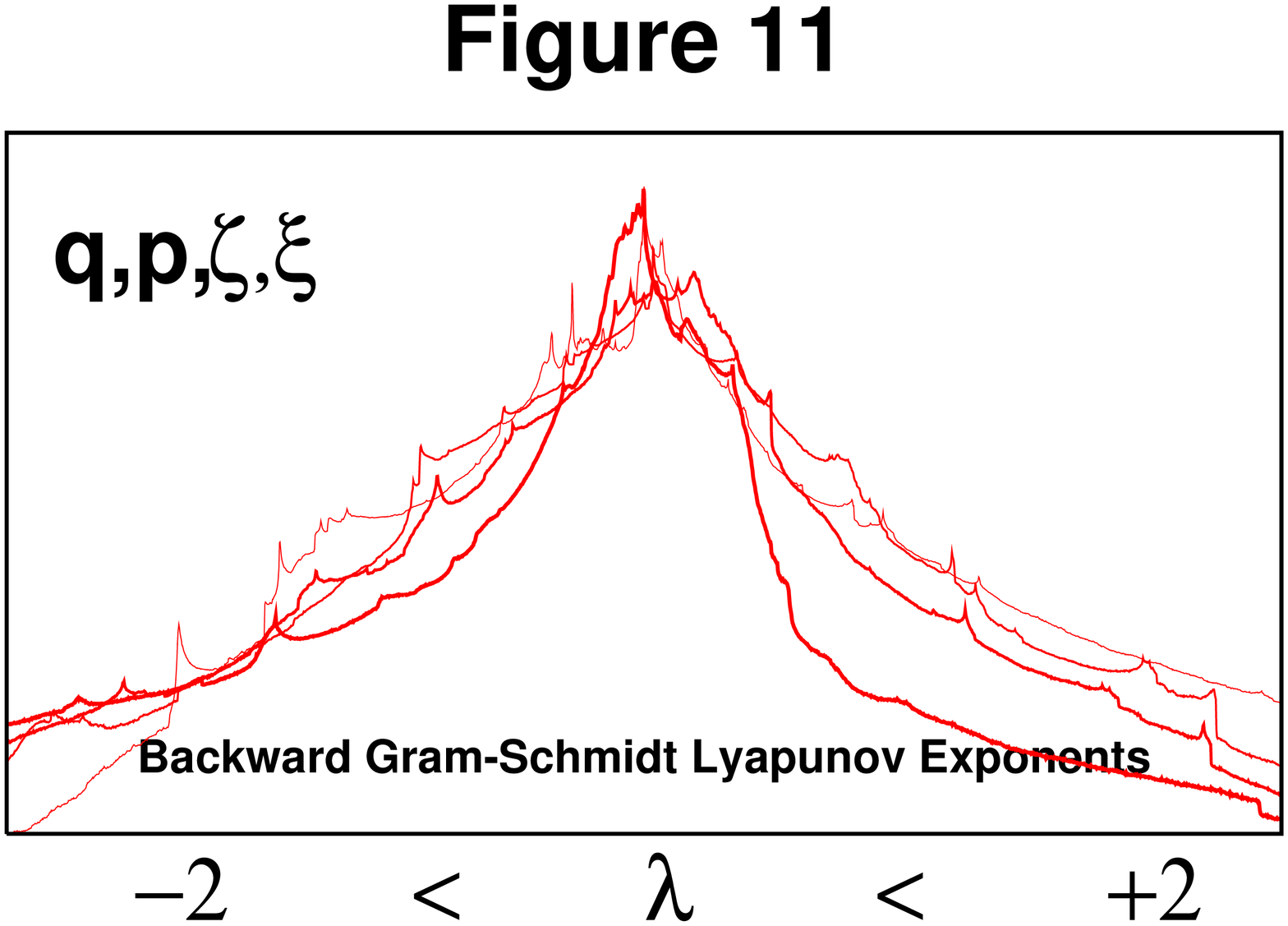}
\vspace{.5in}
\caption{
Probability density for the four backward-in-time Lyapunov exponents,
$\{ \ \lambda^b_{ii}(t) \ \}$.  Notice that the probability distribution
for the most positive of the backward exponents, $\lambda^b_1 = +0.411$
(widest line) matches that for the covariant Lyapunov exponent
$\lambda^c_{44}(t)$ shown in Figure 12.  The logarithmic
ordinate scale covers two decades with data from the central
two billion out of four billion fourth-order Runge-Kutta timesteps of 0.001
each.
}
\end{figure}

\begin{figure}
\includegraphics[height=7cm,width=4cm,angle=-90]{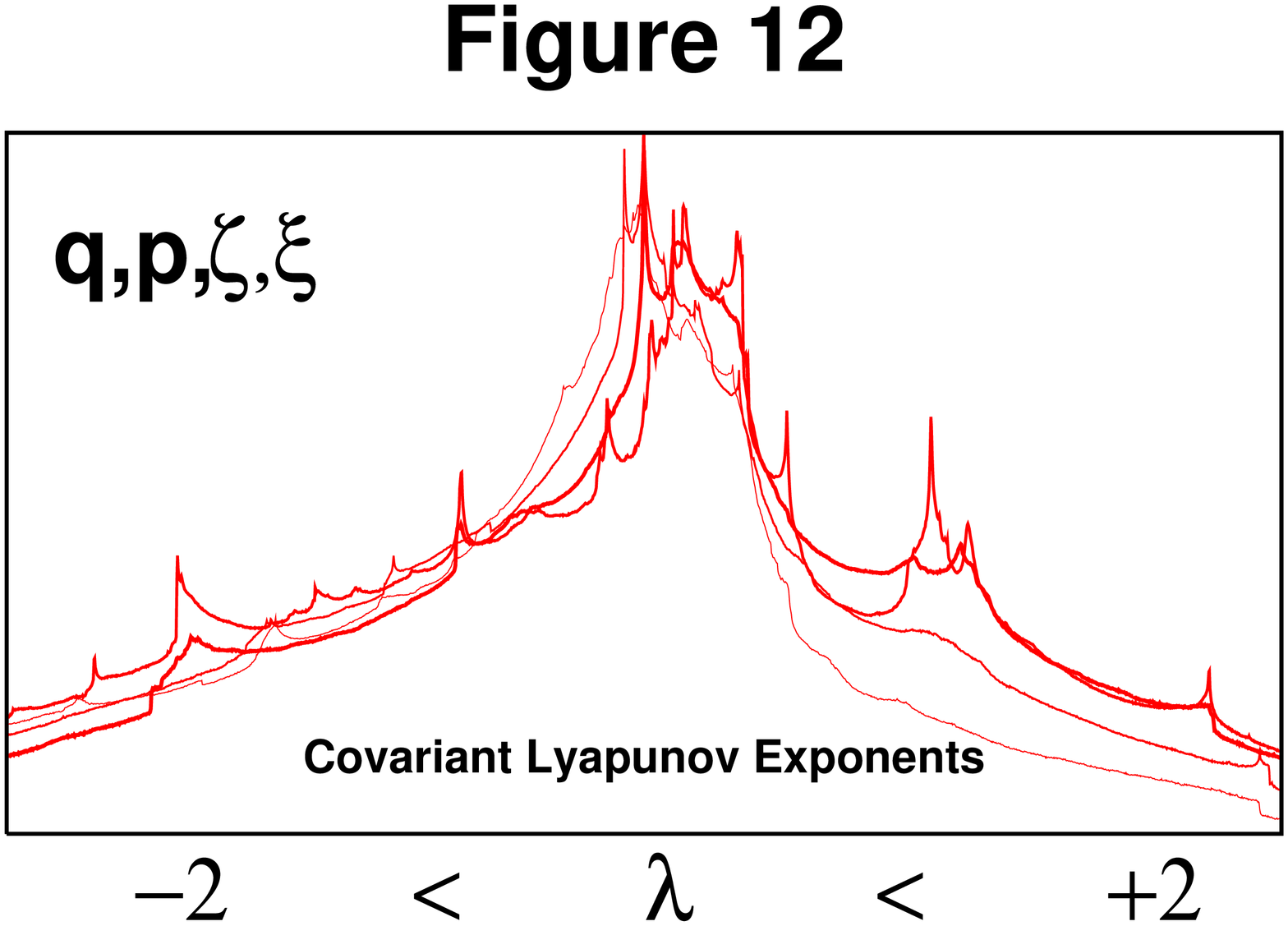}
\vspace{.5in}
\caption{
Probability density for the four covariant Lyapunov exponents,
$\{ \ \lambda^c_{ii}(t) \ \}$. The largest and smallest of these exponents
match the largest forward and the largest backward of the Gram-Schmidt
exponents. The covariant vectors are identical forward or backward in
time, with their exponents simply changing sign.  The logarithmic ordinate
covers two decades with data from the central
two billion out of four billion fourth-order Runge-Kutta timesteps of 0.001
each.
}
\end{figure}

\section{Perspectives}

The covariant phase-space vectors, illustrated here for oscillators, are
said to have two advantages:
[1] they display the time-symmetry of the underlying equations of motion and
[2] they provide results which are ``norm-independent".  The Gram-Schmidt
vectors differ, reflecting the past and steadfastly ignorant of the future.
Consider
a purely-Hamiltonian situation, the sudden inelastic collision of two
rapidly-moving blocks, converting ordered kinetic energy to heat.  The
Gram-Schmidt vectors turn out to be more localized in the forward
direction of time than in the (completely unphysical) reversed direction,
where the particle trajectories have been stored\cite{b29}.  The phase-space
directions corresponding to maximum growth and decay (easy to calculate
by singular value decomposition of the symmetrized $D$ matrix and relatively
smooth in phase space, rather than multifractal) show the same time
symmetry as do the covariant vectors.  This time symmetry seems unhelpful
for describing the irreversibility of this purely-Hamiltonian shockwave
process.  We like the time asymmetry of the Gram-Schmidt vectors,
reflecting as they do the past-based nature of the Second Law of
Thermodynamics.

Despite the norm-independence of the Lyapunov vectors it is apparent
that ``local'' (in phase-space or in  time) Lyapunov exponents depend
on the chosen coordinate system.  Even the simple harmonic oscillator
considered here shows coordinate-dependent local exponents.

The time symmetry-breaking associated with time-reversible dissipative
systems (like the three-dimensional and four-dimensional oscillator
models studied here) prevents backward analyses unless trajectories are
stored.  Lyapunov instability helps {\it attract} a repellor state
(violating the Second Law) to the Law-abiding strange attractor.  Like
the overall trajectory, the Gram-Schmidt vectors are themselves unstable
to symmetry breaking.  A reversed trajectory calculation will not also
reverse the Gram-Schmidt vectors unless those vectors too are stored
from a forward trajectory.  The covariant vectors are the same going
forward and backward in time.  The symmetry breaking which naturally
occurs can only be avoided by repeating a trajectory, making it possible
to find the covariant solution.

Despite the double-precision accuracy of Runge-Kutta solutions it is
difficult to validate computer programs for dissipative chaotic systems
precisely due to their Lyapunov instability.  For this reason the
squared projections of the offset vectors and the exponent histograms
are good choices for validation.  The details of specific program results
necessarily vary and reflect the fractal nature of the phase-space
singularities describing the inevitable past and future bifurcations.
The fractal nature of these singular points frustrates any attempt to
gain accuracy through mesh refinement.  Both the covariant and the Gram-Schmidt
exponents share this fractal nature, but differently.  The Gram-Schmidt
vectors are analogous to passengers' reactions to a curvy highway while the
covariant view is that of a stationary pedestrian observer.

\section{Acknowledgments}

We thank Mauricio Romero-Bastida and Antonio Politi for useful conversations
at a workshop organized by Thomas Gilbert and Dave Sanders, held
in Cuernavaca, Mexico earlier this year.  Rainer Klages, Pavel Kuptsov,
Roger Samelson, Franz Waldner, and Christopher Wolfe generously
provided some comments and useful literature references.  We specially thank
Lakshmi Narayanan for stimulating this work by requesting a Second Edition
of our book on Time Reversibility, and Stefano Ruffo, for expediting
publication in Communications in Nonlinear Science and Numerical Simulation.


\begin{thebibliography}{99}

\bibitem{b1}  W. G. Hoover, {\it Time Reversibility, Computer Simulation,
              and Chaos} (World Scientific, Singapore, 2001).

\bibitem{b2}  W. G. Hoover, {\it Computational Statistical Mechanics}
              (Elsevier, Amsterdam, 1991, and available free at
              http://www.williamhoover.info).

\bibitem{b3}  W. G. Hoover and W. T. Ashurst, ``Nonequilibrium Molecular
              Dynamics'', Theoretical Chemistry {\bf 1}, 1-51 (Academic
              Press, New York, 1975).

\bibitem{b4}  W. G. Hoover, C. G. Hoover, and J. Petravic, ``Simulation
              of Two- and Three-Dimensional Dense-Fluid Shear Flows
              {\it via} Nonequilibrium Molecular Dynamics: Comparison of
              Time-and-Space-Averaged Stresses from Homogeneous Doll's
              and Sllod Shear Algorithms with Those from Boundary-Driven
              Shear'', Physical Review E {\bf 78}, 046701 (2008). 

\bibitem{b5}  J. D. Farmer, E. Ott, and J. A. Yorke, ``The Dimension of
              Chaotic Attractors'', Physica {\bf 7D}, 153-180 (1983).

\bibitem{b6}  S. D. Stoddard and J. Ford, ``Numerical Experiments on the
              Stochastic Behavior of a Lennard-Jones Gas System'',
              Physical Review A {\bf 8}, 1504-1512 (1973).

\bibitem{b7}  I. Shimada and T. Nagashima, ``A Numerical Approach to
              Ergodic Problem of Dissipative Dynamical Systems'',
              Progress of Theoretical Physics {\bf 61}, 1605-1616 (1979).

\bibitem{b8}  G. Benettin, L. Galgani, and J. M. Strelcyn, ``Kolmogorov
              Entropy and Numerical Experiments'',
              Physical Review A {\bf 14}, 2338-2345 (1976).

\bibitem{b9}  E. N. Lorenz, ``A Study of the Predictability of a 28-Variable
              Atmospheric Model'', Tellus {\bf 17}, 321-333 (1965).

\bibitem{b10} E. N. Lorenz, ``Lyapunov Numbers and the Local Structure of
              Attractors'', Physica {\bf 17D}, 279-294 (1985).

\bibitem{b11} Almost all of Edward Norton Lorenz' publications can be
              found on an MIT webpage,
\newline
              [ http://eapsweb.mit.edu/research/Lorenz/publications/htm ].

\bibitem{b12} J. P. Eckmann and D. Ruelle, ``Ergodic Theory of
              Chaos and Strange Attractors'', Reviews of Modern Physics
              {\bf 57}, 617-656 (1985).

\bibitem{b13} A. Trevisan and F. Pancotti, ``Periodic Orbits, Lyapunov
              Vectors, and Singular Vectors in the Lorenz System'', Journal
              of the Atmospheric Sciences {\bf 55}, 390-398 (1998).

\bibitem{b14} C. L. Wolfe, ``Quantifying Linear Disturbance Growth in Periodic
              and Aperiodic Systems'', Ph. D. Dissertation written under the
              Supervision of Roger Samelson, at Oregon State University, 2006.

\bibitem{b15} F. Ginelli, P. Poggi, A. Turchi, H. Chat\'e, R. Livi, and
               A. Politi, ``Characterizing Dynamics with Covariant Lyapunov
               Vectors'', Physical Review Letters {\bf 99}, 130601 (2007).

\bibitem{b16} C. L. Wolfe and R. M. Samelson, ``An Efficient Method for
              Recovering Lyapunov Vectors from Singular Vectors'',
              Tellus {\bf 59A}, 355-366 (2007).

\bibitem{b17} M. Romero-Bastida, D. Paz\'o, J. M. L\'opez, and
              M. A. Rodriguez, ``Structure of Characteristic Lyapunov Vectors
              in Anharmonic Hamiltonian Lattices'', Physical Review E {\bf
              82}, 036205 (2010).

\bibitem{b18}  H. Bosetti, H. A. Posch, Ch. Dellago, and W. G. Hoover,
              ``Time-Reversal Symmetry and Covariant Lyapunov Vectors for the
              Doubly Thermostated Harmonic Oscillator'', Physical Review E
              {\bf 82}, 046218 (2010).

\bibitem{b19} P. V. Kuptsov and U. Parlitz, ``Theory and Computation of
              Covariant Lyapunov Vectors'', arXiv 1105.5228 (Nonlinear
              Sciences, Chaotic Dynamics) (26 May 2011).

\bibitem{b20} W. G. Hoover, C. G. Hoover, and H. A. Posch, ``Lyapunov
              Instability of Pendulums, Chains, and Strings'', Physical
              Review A {\bf 41}, 2999-3005 (1990).

\bibitem{b21} W. G. Hoover and H. A. Posch, ``Direct Measurement of
              Equilibrium and Nonequilibrium Lyapunov Spectra'', Physics
              Letters A {\bf 123}, 227-230 (1987).

\bibitem{b22} I. Goldhirsch, P. L. Sulem, and S. A. Orszag, ``Stability
              and Lyapunov Stability of dynamical Systems: a Differential
              Approach and a Numerical Method'', Physica {\bf 27D},
              311-337 (1987).

\bibitem{b23} W. G. Hoover, Canonical Dynamics: Equilibrium Phase-Space
              Distributions'', Physical Review A {\bf 31}, 1695-1697 (1985).

\bibitem{b24} H. A. Posch, W. G. Hoover, and F. J. Vesely, ``Canonical
              Dynamics of the Nos\'e Oscillator: Stability, Order, and
              Chaos'', Physical Review A {\bf 33}, 4253-4265 (1986).

\bibitem{b25} Wm. G. Hoover, C. G. Hoover, and F. Grond, ``Phase-Space
              Growth Rates, Local Lyapunov Spectra, and Symmetry Breaking for
              Time-Reversible Dissipative Oscillators'', Communications in
              Nonlinear Science and Numerical Simulation {\bf 13}, 1180-1193
              (2008).

\bibitem{b26} Wm. G. Hoover, C. G. Hoover, and H. A. Posch, ``Dynamical
              Instabilities, Manifolds, and Local Lyapunov Spectra Far From
              Equilibrium'', Computational Methods in Science and Technology
              {\bf 7}, 55-65 (2001).

\bibitem{b27} Wm. G. Hoover, C. G. Hoover, H. A. Posch, and J. A. Codelli,
              ``The Second Law of Thermodynamics and Multifractal Distribution
              Functions: Bin Counting, Pair Correlations, and the [Definite
              Failure of the] Kaplan-Yorke Conjecture'', Communications in
              Nonlinear Science and Numerical Simulation {\bf 12}, 214-231
              (2005).

\bibitem{b28} Wm. G. Hoover and B. L. Holian, ``Kinetic Moments Method for
              the Canonical Ensemble Distribution'', Physics Letters A
              {\bf 211}, 253-257 (1996).

\bibitem{b29} Wm. G. Hoover and C. G. Hoover, ``Three Lectures: NEMD,
              SPAM, and Shockwaves", presented at the Granada Seminar on
              the Foundations of Nonequilibrium Statistical Physics, 13-17
              September, 2010: arXiv:1008.4947.






\end{thebibliography}
\end{document}